\documentclass[%
 reprint,
superscriptaddress,
 amsmath,amssymb,
 aps,
pra,
]{revtex4-2}
\usepackage{graphicx}
\usepackage{placeins}
\usepackage{siunitx}
\usepackage{comment}
\usepackage{natbib}
\usepackage{xcolor}
\usepackage{color}
\usepackage{amssymb}
\usepackage{bm}
\usepackage{comment}
\usepackage{hyperref}
\hypersetup{
    colorlinks=true,
    citecolor=blue,
    linkcolor=blue,
    urlcolor=blue
    }
\makeatletter
\def\NAT@def@citea{\def\@citea{\NAT@separator}}
\makeatother
\newcommand{\bos}[1]{\bm{#1}}

\usepackage{soul}

\def\epsi{\varepsilon}
\def\vphi{\varphi}
\def\eparam{\varepsilon}

\def\tr{^\text{T}}

\def\Eh{E_\text{h}}

\usepackage{soul}

\begin{document}
\preprint{APS/123-QED}

\title{%
Lower bounds on par with upper bounds for few-electron atomic energies
}

\author{Miklos Ronto}
\email{miklos.ronto@weizmann.ac.il}
\affiliation{Chemical and Biological Physics Department, Weizmann Institute of Science, 76100 Rehovot, Israel}
\author{Peter Jeszenszki}
\email{jeszenszki.peter@ttk.elte.hu}
\affiliation{ELTE, Eötvös Loránd University, Institute of Chemistry, Pázmány Péter sétány 1/A, Budapest, H-1117, Hungary}
\author{Edit M{\'a}tyus}
\email{edit.matyus@ttk.elte.hu}
\affiliation{ELTE, Eötvös Loránd University, Institute of Chemistry, Pázmány Péter sétány 1/A, Budapest, H-1117, Hungary}
\author{Eli Pollak}
\email{eli.pollak@weizmann.ac.il}
\affiliation{Chemical and Biological Physics Department, Weizmann Institute of Science, 76100 Rehovot, Israel}

\date{\today}

\begin{abstract}
The development of computational resources has made it possible to determine upper bounds for atomic and molecular energies with high precision. 
Yet, error bounds to the computed energies have been available only as estimates.
In this paper,
the Pollak--Martinazzo lower bound theory, in conjunction with correlated Gaussian basis sets, is elaborated and implemented to provide sub-parts-per-million convergence of the ground and excited state energies for the He, Li, and Be atoms.  
The quality of the lower bounds is comparable to that of the upper bounds obtained from the Ritz method. These results exemplify the power of lower bounds to provide tight estimates of atomic energies.
    
\end{abstract}

\maketitle

\section{Introduction}
A century ago, the development of quantum theory was motivated, among others, by the stability of atoms
and molecules. Schrödinger’s Coulomb Hamiltonian for the hydrogen
atom has a finite, lowest energy eigenvalue, {i.e.,}
quantum theory correctly predicted its stability. 
Regarding poly-electronic and poly-atomic systems, the analytic solution is unknown, but 
it has been demonstrated by formal tools that the many-particle Coulomb
Hamiltonian is bounded from below~\cite{Lieb2005,simonSpectralTheoryMathematical2007}.
In this sense, formal
lower bound theory played an essential role in showing
that non-relativistic quantum theory was qualitatively
correct.

The evaluation of the ground and excited state energy levels of the Hamiltonian has been of central importance during the course of the practical application of quantum theory to molecular physics and chemistry.
The Schrödinger equation of atomic and molecular systems has been solved by various numerical techniques; the most accurate energy values have been obtained by variational methods. 

Variational methods are based on the variational principle, formulated for an energy upper bound, and provide systematic numerical means to converge from above to the (unknown) exact energy (and the corresponding wave function) by using computer power.

In spite of the essential role lower bounds played in the formal theory, they were rarely used as practical computational tools and for good reason.
Computed lower bounds have been orders of magnitude less accurate than the upper bound; thus, the computational effort was concentrated on converging the upper bound. The convergence rate of the upper bound has been used to estimate the exact non-relativistic energy (within some estimated energy interval). Such an extrapolation is approximate and may fail. The energy
uncertainties derived from basis set extrapolation have sometimes turned out to be overly optimistic, making conclusions based on estimated error bars to the computed energies unreliable.

The present work aims to turn the formal lower bound theory into a practical computational tool that provides an energy lower bound converging to the (unknown) exact energy value from below at a rate comparable to the upper bound. 
Thereby, it becomes possible to compute and systematically narrow the energy interval within which the exact non-relativistic energy resides. 
This procedure allows us to take the first step towards ensure{computing} error intervals, instead of estimating them. A computed error interval to the computed atomic (or molecular) energy is necessary for a good comparison with experimental data, when we aim to test and further develop the fundamental theory of atomic and molecular matter. 
In this work, we present algorithmic developments and computations for few-electron atoms. Further work is planned to generalize the procedure for molecular energies.

Several lower bound methods have been introduced based on the Temple~\cite{Temple1928} and the Weinstein~\cite{Weinstein1934} approaches. The Weinstein lower bound was further elaborated and generalized by Stevenson and Kato~\cite{Stevenson1938a,Stevenson1938b,Kato1949}. Several theoretical~\cite{Kato1950,Temple1952,Delves1972,Kleindienst1976,Harrell1978,Cohen1979,Marmorino2002,Marmorino2004} and practical improvements~\cite{Kleindienst1986,King1995,Marmorino2012} have been developed with respect to the Temple bound. The optimal inclusion intervals introduced by Lehmann~\cite{Lehmann1949,Lehmann1950,Beattie1998} were a significant development in relation to the original Temple bound. Further approaches of lower bound methods are based on bracketing functions~\cite{Lowdin1964,Lowdin1965a,Lowdin1965b,Scrinzi1992,Szabados2014,Toth2015} and on the method of intermediate operators~\cite{Weinstein1972,Aronszajn1951}. Lower bounds are also of importance in the context of physical properties of few-electron atoms such as oscillator strengths~\cite{Sims1,Sims2,Sims3,Sims4,Marmorino2020}. 

In the past few years, a novel class of lower bound methods~\cite{Pollak2019a,Pollak2019b} based on the Lánczos construction of basis sets has been proposed. 
A self-consistent lower bound theory (SCLBT)~\cite{Martinazzo2020a,Pollak2020a} was developed and successfully applied to quartic~\cite{Pollak2020a} and double-well~\cite{PollakRonto2020} potentials as well as lattice models~\cite{Martinazzo2020a,PollakRonto2021}; however, these methods are typically not applicable for Coulomb-interacting systems due to the divergence of matrix elements of cubic and higher powers of the Coulombic Hamiltonian, unless one can devise basis sets, which like the true eigenfunctions, prevent such divergence and still the basis is in principle complete.

While a tight Temple lower bound was computed for the helium atom~\cite{Nakashima2008}, the quality of this lower bound is several orders of magnitude worse than the corresponding upper bound. As alternatives to Temple's approach, expanding on Aronszajn's work~\cite{Aronszajn1951}, Bazely~\cite{Bazley1960,Bazley1961} and later Bazely and Fox~\cite{Bazley1964} obtained lower bounds using intermediate Hamiltonians by introducing a special choice for the finite-dimensional space used to represent the Hamiltonian operator. This class of methods has been further elaborated and applied to Coulombic systems as well as other potentials~\cite{Gay1964,Miller1965,Weinhold1968,Hill1980}, most recently by Marmorino~\cite{Marmorino2008,Marmorino2013}. Lower bounds to the energy eigenvalues of the helium atom have been computed using further strategies~\cite{Marmorino2000,Marmorino2011} and an energy lower bound to the ground state of the lithium atom was computed using Lehmann's method~\cite{luchow}. However, none of these studies resulted in lower bounds with comparable accuracy to those obtained with the Ritz variational method.

Most recently, a different approach has been introduced by Pollak and Martinazzo applicable for Coulombic potentials, which was successfully used to compute lower bounds to the energy levels of hydrogen~\cite{Pollak2021}, and the two-electron helium and the three-electron lithium atoms~\cite{Ireland2021}.
This work presents a further development and application of the method based on the use of explicitly correlated Gaussian basis sets. We report algorithmic and computational strategies and present numerical results for lower bounds to the (ground and excited state) energies of the helium, lithium, and beryllium atoms with a relative precision comparable to the corresponding upper bound obtained in the same series of computations.

\section{Lower bound theory}

Schrödinger's formulation of the non-relativistic Hamiltonian 
of atoms with a fixed nucleus of  charge number $Z$ is written 
in Hartree atomic units  as
\begin{align}
    H
    =
    -\frac{1}{2}
    \sum_{i=1}^N
      \Delta_{_i}
    +\sum^N_{i=1}\sum^N_{j>i}\frac{1}{r_{ij}}
    -\sum_{i=1}^N\frac{Z}{r_i}\label{Hamil} \; . 
\end{align}
with $r_i$ denoting the distance of the $i$th electron from the nucleus,  $r_{ij}$ denoting the distance of the $i$th electron from the $j$th one, and $\Delta_{_i}$ is the kinetic energy operator for the $i$th electron. The nucleus is assumed to be stationary, with infinite mass, located at the origin of the spatial coordinate system.
A central branch of molecular physics revolves about the computation of 
stationary states of $H$ by (numerical) solution of the eigenvalue equation
\begin{align}
  H \psi_n = \epsi_n \psi_n \; ,\quad n=1,2,\ldots\ .
\end{align}
According to the Ritz--Macdonald variational principle~\cite{Ritz1909,MacDonald1933}, 
the energy functional 
\begin{align}
  \lambda_n
  =
  \frac{
    \langle \vphi_n|H|\vphi_n\rangle
  }{%
    \langle \vphi_n|\vphi_n\rangle
  }\ge \epsi_n,
  \label{eq:genvar}
\end{align}
provides an upper bound to the exact energy $\varepsilon_n$ for an appropriate $\vphi_n$ trial function. For a linear parametrization of the trial function in terms of ``primitive'' basis functions, $f_i$,  $\vphi_n=\sum_{i=1}^{L}c_{ni} f_i$, the minimization problem is turned 
into a matrix eigenvalue equation
\begin{align}
  \bos{H} \bos{c}_n
  =
  \lambda_n^{(L)} \bos{S} \bos{c}_n
  \; , \label{eq3} 
\end{align}
where the $\bos{H}$  Hamiltonian and $\bos{S}$ overlap matrices are calculated using  the basis functions. The matrix elements are $H_{ij}=\langle f_i |H| f_j \rangle$ and $S_{ij}=\langle f_i|f_j\rangle$.

A central difficulty in computing lower bounds to Coulombic systems using Lánczos basis sets or more generally the Krylov algorithm \cite{Saad2003} is due to the fact that with most basis sets, $H^2$ is the highest power of the Coulomb Hamiltonian that can be handled; powers greater than 2 usually diverge. The expectation value of $H^2$, $\langle H^2\rangle_n =\langle \vphi_n|H^2|\vphi_n\rangle$, and the corresponding variance,
$\sigma_n^2 = \sqrt{\langle H^2\rangle_n - \langle H\rangle_n^2}$, can be computed and used in relation with several lower bound theories. However, until recently, all lower bound theories returned numerical values that were several orders of magnitude less accurate than the upper bound obtained in a similar computational setup; hence, the practical utility of the computed lower bounds remained limited.

The recently formulated  Pollak--Martinazzo (PM) lower bound theory addresses this problem by constructing a special matrix used in conjunction with the Cauchy interlacing theorem. According to the interlacing theorem, which can be derived from the Courant--Fisher theorem~\cite{Hwang2004}, if the eigenvalues of an $n\times n$ Hermitian matrix $A$ are given in ascending order as  $a_1\leq a_{2}\leq\ldots\leq a_{n-1}\leq a_{n}$, and the eigenvalues of its $(n-1)\times(n-1)$ principal submatrix $B$ are
$b_1\leq b_{2}\leq\ldots\leq b_{n-2}\leq b_{n-1}$, then $a_1\leq b_{1}\leq a_{2}\leq b_{2}\leq\ldots\leq a_{n-1}\leq b_{n-1}\leq a_n$. 
This theorem is used to obtain lower bounds to the eigenvalues of $B$ as follows. The matrix $B$ with dimension $L=n-1$ is substituted with the diagonal Hamiltonian matrix obtained by diagonalizing the $L\times L$ Hamiltonian matrix, Eq.~\eqref{eq3}, with eigenvalues denoted in an ascending order as $\lambda_j^{(L)}$. 

Then, a ``big'' $[(L+1)\times(L+1)]$-dimensional matrix is defined, motivated by the matrix $A$ in the previous paragraph,  as \cite{Pollak2021}
\begin{equation}
{\boldsymbol{K}}_{L}(\eparam )=\left(
\begin{array}{ccccc}
\lambda _{1}^{\left( L\right) } & 0 & \ldots & 0 & \sigma _{1}^{\left(
L\right) } \\
0 & \lambda _{2}^{\left( L\right) } & \ldots & 0 & \sigma _{2}^{\left(
L\right) } \\
\vdots & \vdots & \ddots & 0 & \vdots \\
0 & 0 & \ldots & \lambda _{L}^{\left( L\right) } & \sigma _{L}^{\left(
L\right) } \\
\sigma _{1}^{\left( L\right) } & \sigma _{2}^{\left( L\right) } & \ldots &
\sigma _{L}^{\left( L\right) } & 
\eparam +\displaystyle{\sum_{k=1}^{L}}\frac{\left(
\sigma _{k}^{\left( L\right) }\right) ^{2}}{\lambda _{k}^{\left( L\right)
}-\eparam }%
\end{array}%
\right) \; ,\   \label{PMmat}
\end{equation}%
where $\lambda_n^{(L)}$ labels the $n$th Ritz eigenvalue and $\sigma_n^{(L)}$ is the associated standard deviation.
We refer to the matrix $\bos{K}_L(\varepsilon)$ as the PM matrix with parameter $\varepsilon$.

By construction, the parameter $\varepsilon$ is an eigenvalue of the PM matrix. The remaining eigenvalues of the matrix are the $L$ solutions of the polynomial equation
\begin{equation}
    1=\sum_{k=1}^{L}\frac{\sigma _{k}^{2}}{\left( \lambda _{k}-\varepsilon
\right) \left( x-\lambda _{k}\right) }\; .  \label{polynom1}
\end{equation} 
They are denoted in ascending order as $x_j (j=1,\ldots,L)$ and have the important property that $x_j\left(\varepsilon\right)$ is a monotonically increasing function of the parameter $\varepsilon$.

Suppose that we choose $\varepsilon$ to equal the unknown ground state energy denoted as $\varepsilon_1$. According to Cauchy's interlacing theorem, the  eigenvalues $x_k$ of the  $\boldsymbol{K}_{L}(\varepsilon)$ matrix are interlaced by the Ritz eigenvalues $\lambda_k$ as follows: $\varepsilon_1\leq\lambda_1^{(L)}\leq x_1\leq\lambda_2^{(L)}\leq \ldots \leq \lambda_L^{(L)}\leq x_L$. Then, if we have a lower bound for $x_1$ and compute from Eq.~\eqref{polynom1} the value of $\varepsilon$ that would give the same value of $x_1$, then due to the monotonicity property this value of $\varepsilon$ would necessarily be a lower bound to the ground state energy. If the basis set used is ``good'' in the sense that both $\lambda_1^{(L)}$ and $\lambda_2^{(L)}$ are not too far from the exact eigenvalues $\varepsilon_1$ and $\varepsilon_2$, then, barring special circumstances such as described below for the He atom,  one finds that $x_1\geq\varepsilon_2$, so that bounding $x_1$ from below by a lower bound to the excited state energy gives a lower bound to the ground state energy. Since the eigenvalues of the PM matrix are not very sensitive to the precise value of $x_1$ used, this leads to accurate lower bounds, as shown below.

This procedure may then be continued. For example, if $\lambda_3^{(L)}$ is also not too far from the exact eigenvalue $\varepsilon_3$ then $x_2$ will be larger than $\varepsilon_3$, so that replacing it with a lower bound to $\varepsilon_3$ and finding the two lowest eigenvalues of the PM equation, yields lower bounds to the ground and first excited state energies. This procedure may then be continued for the next excited state, etc.

\section{Computational setup with an explicitly correlated Gaussian basis} 
Explicitly correlated Gaussian (ECG) functions~\cite{Boys1960,Singer1960,Jeziorski1979,Suzuki1998,Mitroy2013}
are commonly used as a spatial basis for atomic and
molecular problems; however, unlike orthogonal
polynomials, they do not provide uniform coverage of space by simply increasing the polynomial order. ECGs can be powerfully used in relation with parametrization by optimization (minimization) of some appropriate target functional. (Regarding nodeless harmonic
oscillator functions used as a basis, see Ref.~\cite{Suzuki1998}.) The parametrization of ECGs with respect to minimization of the energy functional is a powerful means of obtaining
and systematically improving energy upper bounds. While ECGs fail to satisfy the cusp condition~\cite{JeIrFeMa22}, they have general, analytic $N$-particle integrals for most physically relevant operators, which can also be generalized for molecular computations.

In this work, trial functions corresponding to the $S$ ground-state symmetries of the helium, lithium, and beryllium atoms are expressed as anti-symmetrized ($\mathcal{A}$) products of $\phi$ spatial and $\chi$ spin functions
\begin{align}
  f_i(\bos{r},\bos{\sigma}) 
  = 
  {\mathcal{A}} \lbrace%
    \phi_{L,M_L}(\bos{r},\bos{A}_i)
    \chi_{S,M_S}(\bos{\sigma},\vartheta_i)   
  \rbrace \; .  \label{eq:Anstaz}
\end{align}
$\chi_{S,M_S}(\bos{\sigma},\vartheta_i)$ corresponds to the two-, three-, and four-electron spin functions coupled to spin states with total spin quantum numbers $(S,M_S)=(1,0)$ for helium and beryllium and to $(2,0)$ for lithium. The total spin functions for lithium and beryllium  correspond to a two-dimensional spin space, which is parametrized by one free parameter ($\theta_i$)~\cite{Suzuki1998}.
We used ECGs as spatial basis functions corresponding to $(L,M_L)=(0,0)$ orbital momentum quantum numbers (suppressed in the rest of the paper),
\begin{align}
  \phi(\bos{r},\bos{A}_i)
  =
  e^{-\bos{r}\tr (\bos{A}_i\otimes \bos{I}_3) \bos{r}} \;  ,  \label{eq:ECG} 
\end{align}
centered at the origin (where the nucleus is fixed) and $\bos{r}\in\mathbb{R}^{3N}$ collects the electronic coordinates. The $\bos{A}_i\in\mathbb{R}^{N \times N}$ positive-definite, symmetric matrix 
determines the width of the Gaussian and the correlation length of the particles, and is determined by optimization of some appropriate target function.

All computations were performed using a computer program named QUANTEN (QUANTum mechanical description of electrons and atomic nuclei) and developed by the Budapest group. QUANTEN has a (stochastic and deterministic) variational engine and an extensive ECG library with recent applications including non-adiabatic, pre-Born--Oppenheimer, perturbative, and variational relativistic computations \cite{JeIrFeMa22,Ma18he2p,FeMa19HH,Ma19rev,FeMa19EF,FeKoMa20,JeFeMa21,JeFeMa22,FeJeMa22,FeJeMa22b}. 
It can be efficiently run with double precision arithmetic, but a quadruple precision mode is also available.
The first implementation of the $H^2$ integrals and assembling the variance computation in QUANTEN, which was recently reported for two- and three-electron atoms~\cite{Ireland_TDK,Ireland2021}, is further developed  and extended to four (and, in general, $N$) particle systems in the present work.

\subsection{Strategy for converging the PM lower bound to the energy}
Atomic PM lower bounds have been reported for the helium and lithium ground states using the computational setup described above~\cite{Ireland2021}. Although, the PM bounds were tighter than the Weinstein, Temple, or Lehmann bounds obtained with the same basis set~\cite{Ireland2021}, even the best PM bound was (at least) three orders of magnitude less precise (a relative precision of 0.55~ppm for helium and 4.0~ppm for lithium was achieved), than the corresponding upper bound (with a relative precision of 0.000\ 17~ppm for helium and 0.002~ppm for lithium). 

The natural question arises: How can we improve the convergence of the PM bounds? The plausible idea of fine-tuning the ECG basis parametrization based on a simple PM energy ensure{maximization} condition was found to be impractical in Ref.~\cite{Ireland2021}. If one is not careful, then a simple minded application of the PM method may lead to energy values which are {\it higher} rather than {\it lower} than the true eigenvalue under study.

To better understand the conditions for which the PM method leads to lower bounds, it is necessary to consider that all lower bound theories based on the variances of the Hamiltonian are only valid under certain conditions. These conditions are typically connected to the quality of the variances and the $\varepsilon$ parameter. When the upper bounds are ``well behaved,'' in the sense that the $\lambda_j-\varepsilon_j$ distance may be considered as small, one may expand the PM equation, Eq.~\eqref{polynom1}, to leading order in the distance to find \cite{Pollak2021}
\begin{equation}
 x_j(\varepsilon)-\varepsilon_{j+1}
 \simeq 
 \lambda_{j+1}-\varepsilon_{j+1}
 -\frac{\sigma_{j+1}^2}{\sigma_j^2}(\lambda_j-\varepsilon_j)\geq0 \; . 
 \label{eq:expand}
\end{equation}
This relation implies that the left-hand  side of the equation will be positive if the ratio of the variances of the $(j+1)$th to the $j$th state is sufficiently small. This suggests that if we are interested, for example, in a high-quality lower bound to the ground-state energy, and we already have a fairly good description of the ground-state upper bound, then we should continue improving not the ground but the first-excited state's description, {i.e.,} continue with the minimization of the first-excited-state energy and associated reduction of its variance.
This is the core idea for the computational developments presented in this paper. Furthermore, Eq.~\eqref{eq:expand}
will also be used to rationalize some further observations regarding the numerical results [sensitivity of the computed lower bounds to the $\varepsilon$ parameter of the PM matrix, Eq.~\eqref{PMmat}, and possible failure of obtaining a lower bound].

The implementation of the core idea, {i.e.,} improvement of the description of excited states to have a better lower bound for the ground state, was not readily available in the existing computational setup. Although ECG basis sets generated based on the energy minimization condition for a selected state provide a very compact representation, they do not guarantee a high-quality description of other states (unlike a set of orthogonal polynomials, for which increasing the number of functions, {i.e.,} the polynomial order, automatically ensures more complete coverage of the space, and, hence, improved convergence of excited-state energies).

\subsubsection{Implementation of the multi-state energy minimization strategy in an ECG-based procedure}
A usual energy minimization procedure, {e.g.,} for the ground state, is initiated by 
random basis generation and selection~\cite{Suzuki1998},
which is followed by repeated refinement cycles of the already existing basis set, for which we use the Powell method~\cite{Powell2006}. Both steps are based on the energy minimization condition (and the variational principle for Hamiltonians bounded from below). 

The same procedure can be repeated for the first- ($n$th) excited state (even long-lived states embedded in the continuum~\cite{FeMa19EF} in combination with a stabilization-like procedure). In this fashion, separate near-optimal basis sets for separate states can be straightforwardly generated.
One could then try and merge the basis sets optimized for the ground and for the first-excited states, but this procedure would result in a gigantic basis and, more importantly, near-linear dependency problems in the finite precision arithmetic used for the computations.

Instead, we have implemented a multi-state procedure in a single computation as follows.
The usual basis generation and refinement using the energy minimization condition for the ground state is implemented up to a certain number of basis functions. This number is determined based on the convergence of the Ritz ground-state energy. This results in the first ``block'' of our basis set.
The computation is then continued with the generation and refinement of additional basis functions (second block of the basis set), for which the energy minimization condition for the \emph{second} state (first-excited state) was implemented. 
We have regularly refined (using the Powell method) the entire basis set, one function after the other, by using the energy minimization condition for the ground-state energy for functions belonging to the first basis block,
and the energy minimization condition for the second state for functions belonging to the second basis block.

The repeated full-basis refinement cycles allow us to relax functions in the first block (optimized to the ground state) while the ground-state energy is also (partly) described by the second-block functions (optimized to the first-excited state). Therefore, small deviations from a monotonic decrease of the energy may occur upon enlargement of the basis set. For sufficiently large basis sets and with further, extended optimizations these small deviations from monotonicity can be smoothed out.

By construction, the procedure generates a basis set which is (near) optimal for both the ground and the first-excited states, and the linear--dependency problem is automatically avoided (a new basis function that would have a too large overlap with the existing basis set is discarded or ``weighted down'' with a ``penalty'' correction to the value of the energy functional).  
Furthermore, the procedure can be straightforwardly extended to additional states, and thus, applicable also beyond the ground state.

\subsubsection{Numerical demonstration of the multi-state optimization strategy for lithium and beryllium}
\begin{figure}[t]
  \includegraphics[width=0.5\textwidth]{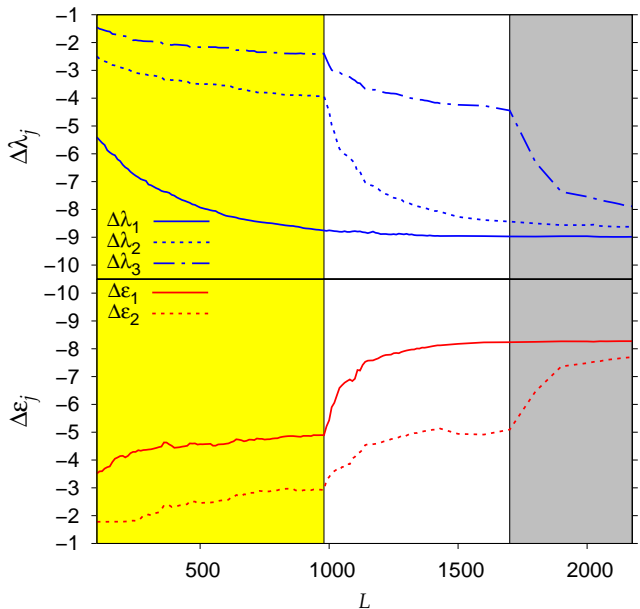}
  \caption{\label{fig:Li} %
    Upper and lower bound gaps [Eq.~\eqref{eq10}] for the lithium atom with respect 
    to the $\varepsilon_{\mathrm{ref},j}$ reference energies taken from Ref.~\cite{WaYaQiDr12}. 
    The first, second, and third basis blocks with $L\in [1,980], (980,1700],$ and $(1700,2175]$  basis indexes, in short [1:980:1700:2175], were optimized according to minimization of the ground (yellow), the first- (white), and the second-excited state (gray) energies. (See also Table~\ref{Litable}.)
  }
\end{figure}

The computational strategy described above has been implemented in QUANTEN. It is highlighted for the case of lithium in Fig.~\ref{fig:Li} (see also Table~\ref{Litable}). 
The computations are more expensive for beryllium, so for this case, we  report only the final results (Tables~\ref{Betable} and~\ref{Betable2}). 
Unexpectedly, helium turned out to be a very special case, for which the strategy does not work (the condition $x_1\geq\varepsilon_2$ fails), and this  can be rationalized on the basis of Eq.~\eqref{eq:expand} as explained in the last paragraphs of this section.

As a measure of the quality of the lower-bound energy for a given basis set, we compare 
the relative upper and lower bound gaps defined as
\begin{equation}\label{eq10}
  \begin{split}
    \Delta\lambda_{j} 
    =& 
    \log_{10}\left(|\lambda_j-\varepsilon_{\mathrm{ref},j}|/\varepsilon_{\mathrm{ref},j}\right) \\
    \Delta\varepsilon_{j} 
    =& 
    \log_{10}\left(|\varepsilon_{j,-}-\varepsilon_{\mathrm{ref},j}|/\varepsilon_{\mathrm{ref},j}\right),
  \end{split}
\end{equation}
where $\varepsilon_{\mathrm{ref},j}$ is a reference value (expected to be very close to the exact value and available from the literature for the computed examples), $\lambda_j$ and $\varepsilon_{j,-}$ are the computed upper and lower bounds for the $j$th state ($j=1,2,\ldots$), respectively. (We note that $\varepsilon_j^-$ is used in Tables~\ref{Litable}--\ref{Hetable} to label the \emph{estimated} lower bounds used in the PM equation.) If the gap ratio, 
\begin{align}
  \eta_j
  =
  \Delta\varepsilon_{j}/\Delta\lambda_{j} \; ,
  \label{eq:gapratio}
\end{align}
approaches one, we may say that the lower (and upper) bound computation is useful in terms of bracketing the exact energy.

In Fig.~\ref{fig:Li}, showcasing our computation for lithium, the performance of the various energy estimates in the  yellow region is comparable to the best gap ratio achieved in Ref.~\cite{Ireland2021}. Then, we continue with the multi-state optimization procedure. During the generation of the second basis block (white region in the figure), we see a significant improvement for the ground-state lower bound, and the first-excited state lower bound also improves (lower part of the figure), in parallel with the improvement of the first- and second-excited-state upper bounds (upper part of the figure).

As can be seen in the figure, the optimization for one state does not necessarily guarantee the monotonic improvement of the ground and other states; however, any increase in a state energy can be minimized by applying subsequent refinement cycles to the already generated basis set. The figure also shows 
the generation and optimization of a third basis block (gray-shaded area), in which the basis functions are optimized using the energy minimization condition for the second-excited state.

\begin{table}[b]
\caption{Results for the lithium atom. Tabulation of lower- and upper-bound energies, $\varepsilon_{j,-}$ and $\lambda_j$, respectively, in units of $E_\text{h}$, resulting from multi-state optimization with three blocks (with basis sizes [1:980:1700:2175], cf. Fig.~\ref{fig:Li}). The relevant variances are also given in units of $E_\text{h}^2$.
The relative deviation from the $\varepsilon_{\mathrm{ref},j}$ reference energy adapted from Ref.~\cite{WaYaQiDr12} is shown in parentheses in ppb (parts per billion). 
The PM parameter in Eq.~\eqref{PMmat} was $\varepsilon_2^-=\varepsilon_{\text{ref},2}-5\cdot 10^{-8}\ \Eh$ and $\varepsilon_3^-=\varepsilon_{\text{ref},3}-10^{-7}$ $\Eh$ for obtaining the lower bounds for the ground and first excited states, respectively (see also Fig.~\ref{fig:Listab} and corresponding text). 
}
\begin{ruledtabular}
\label{Litable}
\begin{tabular}{@{}llc@{}}
\multicolumn{3}{l}{Ground state} \\
 \hline
 $\varepsilon_{1,-}$ & \num{-7.4780603637500093}  & (5.3) \\   
 $\varepsilon_{\text{ref},1}$ & \num{-7.4780603239068} \cite{WaYaQiDr12}  \\
 $\lambda_1$ & \num{-7.4780603161791626} & (1.0) \\  
\hline
\multicolumn{3}{l}{First excited state} \\ 
\hline
 $\varepsilon_{2,-}$ & \num{-7.3540985691144196}  & (20.1) \\ 
 $\varepsilon_{\text{ref},2}$ &  \num{-7.35409842144437} \cite{WaYaQiDr12}  \\
$\lambda_2$ & \num{-7.3540984039701658}  & (2.4) \\
\hline
\multicolumn{3}{l}{Second excited state} \\ \hline
 $\varepsilon_{\text{ref},3}$ &  \num{-7.31853084599891} \cite{WaYaQiDr12}  \\
$\lambda_3$ & \num{-7.3185307512302507}  & (12.9) \\  
\hline
Variances \\
\hline
$\sigma_1^2$ &\num{0.1589345859529} \\
$\sigma_2^2$&\num{0.225395308656} \\ $\sigma_3^2$&\num{0.2657721219969}\\
\end{tabular}
\end{ruledtabular}
\end{table}

The resulting best upper and lower bound values obtained for the lithium atom corresponding to a total basis size of $L=2175$ are collected in Table~\ref{Litable}. 
While previous PM computations carried out for the lithium atom ground state~\cite{Ireland2021} (with a single basis block) already improved upon the Lehmann bound obtained using a Hylleraas basis~\cite{luchow}, the present PM lower bounds significantly outperform both. The lower bounds are at most one order of magnitude worse than the upper bounds. This may be improved upon if one has a better estimate for the excited state energies as discussed in further detail below. At this point it suffices to say that the values used as estimates for the excited state energies upon implementing the PM equations are rather conservative. 

 As might be expected, the ground-state lower bound is more accurate than the first excited state lower bound and the same ordering of accuracy is true for the upper bounds. The plateauing of the ground-state lower bound for $L\geq 1500$ reflects the plateauing of the ground-state  upper bound. The ground-state  lower bound will improve as the PM eigenvalue $x_1$ converges to the first excited state energy. As seen from Eq.~\eqref{eq:expand}, for this to occur one needs an improvement of the upper bound. Since this does not happen,  the lower bound reaches a plateau value. The same occurs for the first excited state lower bound.

Similarly good results are obtained with multi-state optimization
for the ground- and first-excited states of the beryllium atom (Tables~\ref{Betable} and \ref{Betable2}).
The multi-state optimization strategy was essential to arrive at good lower-bounds also for beryllium.
In this case,  the quality of the ground-state lower bound is comparable to the Ritz upper bound while it is somewhat worse, a factor of $\simeq6$, for the excited state,  This reflects to some extent the lower bound values used for the excited states when implementing the PM equation.

\begin{table}[h]
\caption{%
Results for the beryllium atom: ground state. Tabulation of lower- and upper-bound energies, $\varepsilon_{j,-}$ and $\lambda_j$, respectively, in units of $E_\text{h}$
resulting from multi-state optimization with two blocks (with basis set sizes [1:2000:4500]).  The relevant variances are also given in units of $E_\text{h}^2$.
The relative deviation from the $\varepsilon_{\mathrm{ref},j}$ reference energy given in Ref.~\cite{HoAdBu19} is shown in parentheses in ppb (parts per billion). 
The PM parameter in Eq.~\eqref{PMmat} was 
$\varepsilon_2^-=\varepsilon_{\text{ref},2}- 10^{-7}\ \Eh$
(see also Fig.~\ref{fig:Be} and corresponding text). 
}
\label{Betable}%
\begin{ruledtabular}
\begin{tabular}{@{}llc@{}}
\multicolumn{3}{l}{Ground state} \\ \hline
$\varepsilon_{1,-}$ & \num{-14.66735691691}  & (28) \\ 
$\varepsilon_{\text{ref},1}$ & \num{-14.667356507} \cite{HoAdBu19} & \\
$\lambda_1$ & \num{-14.6673561910} & (22) \\
\hline
\multicolumn{3}{l}{First excited state} \\ \hline
$\varepsilon_{\text{ref},2}$ &  \num{-14.418240364} \cite{HoAdBu19} & \\
$\lambda_2$  &  \num{-14.4182394785436934} & (61) \\ \hline
Variances \\ \hline
$\sigma_1^2$ & \num{0.506745221334597} \\
$\sigma_2^2$& \num{0.698028905530890}\\
\end{tabular}
\end{ruledtabular}
\end{table}

\begin{table}[h]
\caption{
Results for the beryllium atom: first-excited state. Tabulation of lower- and upper-bound energies, $\varepsilon_{j,-}$ and $\lambda_j$, respectively, in units of $E_\text{h}$
resulting from multi-state optimization with three blocks (with basis set sizes [1:2000:4000:4500]).  The relevant variances are also given in units of $E_\text{h}^2$.
The relative deviation from the $\varepsilon_{\mathrm{ref},j}$ reference energy~\cite{HoAdBu19} is shown in parentheses in ppb (parts per billion). 
The PM parameter in Eq.~\eqref{PMmat} was 
$\varepsilon_3^-=\varepsilon_{\text{ref},3}- 10^{-6}\ \Eh$
(see also Fig.~\ref{fig:Be} and corresponding text). 
\label{Betable2}}%
\begin{ruledtabular}
\begin{tabular}{@{}llc@{}}
\multicolumn{3}{l}{First-excited state} \\ \hline
$\varepsilon_{2,-}$ &  \num{-14.4182482052568} & (543) \\
$\varepsilon_{\text{ref},2}$  &  \num{-14.418240364} \cite{HoAdBu19}\\
$\lambda_2$ &  \num{-14.4182389710707710} & (97) \\
 \hline
\multicolumn{3}{l}{Second excited state} \\ \hline
$\varepsilon_{\text{ref},3}$  &  \num{-14.370087938}  \cite{HoAdBu19} \\
$\lambda_3$ &  \num{-14.3700621397689} & (1795) \\ \hline
Variances \\
\hline
$\sigma_2^2$&\num{0.698028905523643} \\ $\sigma_3^2$&\num{5.38990425347825}\\
\end{tabular}
\end{ruledtabular}
\end{table}

\subsection{Stability and sensitivity of the results to the $\varepsilon$ parameter of the PM matrix}
The PM lower-bound  computation (similarly to Temple's bound or other lower-bound methods) requires some knowledge about the higher-energy state(s). This information (estimate) is encoded in the $\varepsilon$ parameter of the PM matrix, [Eq.~\eqref{PMmat}]: for the computation of a lower bound to the $n$th eigenvalue, the $\varepsilon$ value in the PM matrix must be a lower estimate to the $(n+1)$th energy eigenvalue.

The computed lower-bound results (in Tables~\ref{Litable}--\ref{Hetable}) have been reported with a specific $\varepsilon$ value (estimated from a known precise reference value) used in the PM calculation.
The critical reader might comment that obtaining a tight lower bound which is based on knowledge of a different tight lower bound is problematic. Hence, it is necessary to address the ``stability'' of the results with respect to the precise choice of this value. The PM results obtained in previous computations reported in Refs.~\cite{Pollak2021} and~\cite{Ireland2021}  have been found to be relatively insensitive to $\varepsilon$.

\begin{figure}[t]
  \centering
  \includegraphics[width=1.0\columnwidth]{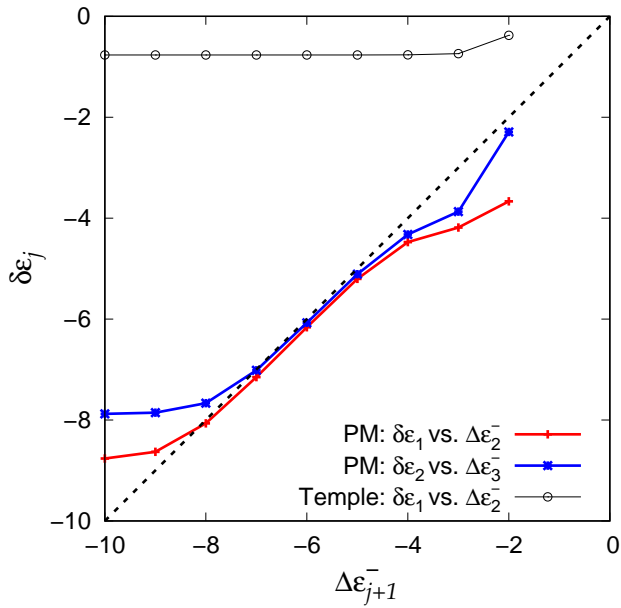}
  \caption{%
    Sensitivity of the PM ground- (red) and first-excited (blue) state lower-bound gap  [Eq.~\eqref{eq:deltaepsij}] for the lithium atom at a basis size of $L=2175$ (cf. Fig.~\ref{fig:Li}) with respect to the precision of the $\varepsilon$ PM parameter used in Eq.~\eqref{PMmat} given in Eq.~\eqref{deps}.  
    The ground-state Temple gap is also shown (black). The ratios of the variances are  $\sigma_2^2/\sigma_1^2=1.42$ and $\sigma_3^2/\sigma_2^2=1.18$.
    \label{fig:Listab}  }
\end{figure}

\begin{figure}[t]
  \centering
  \includegraphics[width=1\columnwidth]{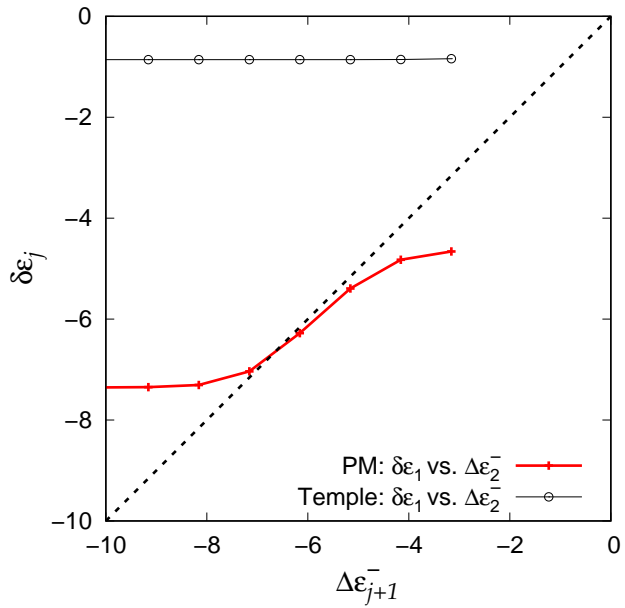}
  \caption{%
    Sensitivity of the PM ground-state lower-bound gap [Eq.~\eqref{eq:deltaepsij}] for the beryllium atom at a basis size of $L=4500$  (2000 states for the ground and 2500 for the first-excited state, (cf. Table~\ref{Betable}) with respect to the precision of the $\varepsilon$ PM parameter used in Eq.~\eqref{PMmat} given in Eq.~\eqref{deps}. The ground-state Temple gap  is also shown (black). The ratio of the variances is
    $\sigma_2^2/\sigma_1^2 = 1.37$.}
    \label{fig:Be} 
\end{figure}

In this work, we repeated the PM computations for the largest basis set results of lithium and beryllium (Tables~\ref{Litable} and \ref{Betable}) using various $\varepsilon$ parameters. 
Figures~\ref{fig:Listab} and~\ref{fig:Be} present the lower bound gap defined with respect to the Ritz eigenvalue 
\begin{align}
  \delta \varepsilon_j
  =
  \log_{10}\left(|\varepsilon_{j,-}-\lambda_j|/\lambda_j\right) \; ,
  \label{eq:deltaepsij}
\end{align}
which is defined analogously to $\Delta \varepsilon_j$ in Eq.~\eqref{eq10}, 
but free  from the knowledge of an ``external'' reference value, $\varepsilon_{\text{ref},j}$. { The gap for the (estimated) $\varepsilon_j^-$ parameter in Eq.~\eqref{PMmat}---which is a lower bound to the respective excited state---is defined with respect to the reference value exactly the same way as the lower bound gap in Eq.~\eqref{eq10}:
\begin{equation}\label{deps}
    \Delta\varepsilon_j^{-}= \log_{10}\left(|\varepsilon_j^{-}-\varepsilon_{\text{ref},j}|/\varepsilon_{\text{ref},j}\right).
\end{equation}
}
In Fig.~\ref{fig:Listab}, the red and blue lines show the ground- and first-excited-state PM lower bound gaps, $\delta\varepsilon_j$, respectively, plotted with respect to the lower-bound gap $\Delta\varepsilon_{j+1}^{-}$,   whereas the black line shows the (orders of magnitude worse) gap for the Temple lower bound, defined as
\begin{align}
  \varepsilon^{\mathrm{Temple}}_{1,-}
  =
  \lambda_1-\frac{\sigma^2_1}{\varepsilon^{-}_2-\lambda_1} \; . 
\end{align}
As can be seen in Figs.~\ref{fig:Listab} and \ref{fig:Be}, the PM lower bounds {\it are} sensitive to the precision of the lower estimate to the $j$th energy ($\varepsilon_j^-= \varepsilon_{\text{ref},j}- \Delta \varepsilon_j^{-}$ is the parameter used in the PM matrix)  while the Temple lower bound is not, due to its poor quality. In the $\Delta \varepsilon_j^{-}$ range used to compute the data reported in Tables~\ref{Litable}--\ref{Betable2}, the functions in Figs.~\ref{fig:Listab} and~\ref{fig:Be} are nearly linear, {i.e.,} the precision of the PM lower bound is determined by the precision of the excited-state estimate used in the PM matrix. This observation can be rationalized on the basis of Eq.~\eqref{eq:expand}. In contrast to the results presented in Refs.~\cite{Pollak2021} and \cite{Ireland2021},  the ratio of variances $\sigma_{j+1}^2/\sigma_j^2$ is of the order of unity, due to the optimization of the excited states, and the accuracy of the $(j+1)$th excited state Ritz eigenvalue is much improved, leading to the linear dependence.

How then does one know the correct value of $\varepsilon$ to be used in the lower bound calculation? The strategy we employed was to use a value that is substantially lower than the accuracy expected from the convergence properties of the relevant Ritz eigenvalue. These are the values reported in Tables ~\ref{Litable}--\ref{Betable2}. The high accuracy of the resulting lower bounds demonstrates that this strategy is robust and that the linear dependence is not really a serious problem.

\subsubsection{The special case of the helium atom}

We applied the multi-state optimization strategy also for the helium atom. When applied naively, the PM equation gave values for the ground-state energy which, in the limit of a large basis set, were {\it larger} than the known ground-state energy. What went awry? 

This is related to the use of a correlated Gaussian basis set rather than an orthogonal polynomial basis. We observed in convergence figures (similar to Fig.~\ref{fig:Li}) 
that the upper bounds (Ritz eigenvalues) to the first- and second-excited states, optimized in the second and third basis blocks, converged faster than the ground-state eigenvalue. 
As may be then reasoned from Eq.~\eqref{eq:expand}, this causes the right-hand side of the equation to be negative, that is, the eigenvalue $x_1$ is no longer greater than the first excited state energy. Using the first excited state energy in the PM equation will then naturally no longer give a lower bound.

It can also be understood that this behavior is unique to the helium atom, which is a two-electron system. The ground state is dominated by a $1s^2$ configuration, the first-excited state is $1s2s$, the second-excited state is $1s3s$, etc. The correlation of the electrons, which is described increasingly more accurately during the course of the variational computation, is less important for excited states, than for the ground state, and hence their Ritz eigenvalues for the excited states converge faster than for the ground state. 

Does this mean that one cannot get meaningful and accurate lower bounds for the He atom using correlated Gaussian basis sets? Not necessarily. If one forces the basis set so that the excited-state eigenvalues are not better than the ground-state level, one may expect the method to work. This is demonstrated in Table~\ref{Hetable}, where using the known excited-state energy values, we can ensure that the accuracy of all three levels is similar. However, this does not answer the question as to how would it be possible, without the knowledge of the numerically exact values, to ensure that the PM equation leads to a lower bound. Fortunately, for larger atoms, the problem does not exist, and as we showed, it is straightforward to obtain high-quality lower bounds for the Li and Be atoms.

\begin{table}[t]
\caption{%
Lower and upper bounds for the He atom energy levels, in $E_\text{h}$, computed in this work for the ground, first-, and second-excited states using $L=510$ ECG basis functions. The relative deviation from the reference energies~\cite{Nakashima2008,Drake} is shown in parentheses in parts per billion (ppb). The PM matrix parameters used in Eq.~\eqref{PMmat} were $\varepsilon_i^-=\varepsilon_{\text{ref},i}-2\cdot 10^{-9}$ $\Eh$ for both $i=2$ and 3. The variances are in units of $E_h^2$.}
\label{Hetable}
\begin{ruledtabular}
\begin{tabular}{@{}llc@{}}
\multicolumn{3}{l}{Ground state} \\
\hline
 $\varepsilon_{1,-}$              & \num{-2.9037243791691136}  & (0.7) \\ 
 $\varepsilon_{\text{ref},1}$ & \num{-2.903724377034119}~\cite{Nakashima2008} & \\
 $\lambda_1$                  & \num{-2.9037243760894203} & (0.3) \\ 
\hline
\multicolumn{3}{l}{First excited state} \\ 
\hline
 $\varepsilon_{2,-}$              & \num{-2.145974047969169}  & (0.9) \\
 $\varepsilon_{\text{ref},2}$ & \num{-2.14597404605359}~\cite{Drake} & \\
 $\lambda_2$                  & \num{-2.1459740450434481}  & (0.5) \\
\hline
\multicolumn{3}{l}{Second excited state} \\ 
\hline
 $\varepsilon_{\text{ref},3}$ & \num{-2.06127198974} \cite{Drake} &  \\
 $\lambda_3$                  & \num{ -2.0612719889718790}         & (0.4) \\ 
 \hline
 Variances \\ \hline
 $\sigma_1^2$ & \num{0.013300746968}\\
  $\sigma_2^2$ & \num{0.1299986192593078}\\
    $\sigma_3^2$ & \num{0.1212166468655}\\
\end{tabular}
\end{ruledtabular}
\end{table}

\section{Summary and discussion}
A multi-state optimization strategy is developed to systematically converge the Pollak--Martinazzo energy lower bound with an explicitly correlated Gaussian basis set. 
Lower bounds to the ground- and first-excited state energies of the lithium and beryllium atoms are computed. The resulting lower bounds are the most precise to date, and their relative precision is comparable to that of the energy upper bound in the same basis.

 In view of  the performance of the multi-state optimization and the PM lower bound theory, the following conclusions can be drawn:
\begin{itemize}
    \item The multi-state optimization of ECG bases provides a systematic and robust improvement of the low-lying eigenvalues.
    \item The optimization of higher lying states does not affect the already converged states adversely.
    \item The optimization of the energy of the $(n+1)$th state improves the quality of the lower bound to the $n$th state.
    \item The PM theory is able to provide lower bounds with ppb relative precision for the energy levels of few-electron
    systems.
\end{itemize}

The presented computational procedure and numerical results are for non-relativistic energies. 
Relativistic and leading-order quantum electrodynamic effects have been traditionally accounted for as perturbative corrections to the non-relativistic energy, {e. g.,} \cite{FeKoMa20}.
The identification of a many-particle relativistic wave equation based on relativistic quantum electrodynamics (QED) is more challenging. Most recently, it became possible (for two particles) to start out from the Bethe--Salpeter QED wave equation, exploit that interactions in atoms and molecules are dominantly instantaneous, and arrive at an eigenvalue equation for a no-pair Dirac--Coulomb--Breit Hamiltonian \cite{MaFeJeMa22}. This Hamiltonian appears to be bounded from below, and robust variational procedures could be developed to compute its eigenvalues, which have an $\alpha$ fine-structure constant dependence that is in agreement with the known $\alpha$ orders of the well-established perturbative procedures \cite{JeFeMa21,JeFeMa22,FeJeMa22,FeJeMa22b}.

This theoretical approach provides variational ensure{relativistic upper bounds} (including also some of the so-called ``non-radiative'' QED corrections of the perturbative framework), and (with further development to many-particle systems), it will be relevant to ask for ensure{relativistic lower bounds} in a spirit similar to this work.

\begin{acknowledgements}
Financial support of the European Research Council through a Starting Grant (No.~851421) is gratefully acknowledged. This work has also been graciously supported by the Ben May Center for Chemical Theory and Computation at the Weizmann Institute of Science.

\end{acknowledgements}
%

\end{document}